\documentclass[showpacs,superscriptaddress,prb,twocolumn]{revtex4}
\usepackage{amssymb}
\usepackage{graphicx}
\newcommand{\equ}[1]{Eq.~(\ref{#1})}
\newcommand{\eqs}[2]{Eqs.~(\ref{#1}) and (\ref{#2})}
\renewcommand{\[}{\begin{equation}}
\renewcommand{\]}{\end{equation}}
\def\bea{\begin{eqnarray}}
\def\eea{\end{eqnarray}}
\def\nn{\nonumber\\}
\def\ee{\varepsilon_\infty}
\def\mm{\mu_\infty}
\def\aa{\alpha_\infty}
\newcommand{\rr}{{\cal R}_\infty}
\newcommand{\zz}{{\cal Z}}

\def\q{{\bf q}}
\def\hq{\hat{\bf q}}
\def\u{{\bf u}}
\def\r{{\bf r}}
\def\E{{\bf E}}
\def\H{{\bf H}}
\def\M{{\bf M}}
\def\B{{\bf B}}
\def\D{{\bf D}}

\def\P{{\bf P}}
\def\today
 {\count10=\year\advance\count10 by -2000 \number\day--\ifcase
  \month \or Jan\or Feb\or Mar\or Apr\or May\or Jun\or
             Jul\or Aug\or Sep\or Oct\or Nov\or Dec\fi--\number\count10}

\def\hour{\count10=\time\count11=\count10
\divide\count10 by 60 \count12=\count10
 \multiply\count12 by 60 \advance\count11 by -\count12\count12=0
\number\count10 :\ifnum\count11 < 10 \number\count12\fi\number\count11}

\begin{document}

\title{Zone-Center Dynamical Matrix in Magnetoelectrics}
 
\author{ R. Resta}
\affiliation{Dipartimento di Fisica, Universit\`a di Trieste, Strada Costiera 11, 34151 Trieste, Italy \\
and DEMOCRITOS National Simulation Center, Istituto Officina dei Materiali (CNR), Trieste.}


\begin{abstract}
In ordinary dielectrics the dynamical matrix at the zone center in general is a nonanalytic function of degree zero in the wavevector $\q$. Its expression (for a crystal of arbitrary symmetry) is well known and is routinely implemented in first principle calculations. The nonanalytic behavior occurs in polar crystals and owes to the coupling of the macroscopic electric field $\E$ to the lattice.
In magnetoelectric crystals both electric and magnetic fields, $\E$ and $\H$, are coupled to the lattice, formally on equal footing. We provide the general expression for the zone center dynamical matrix  in a magnetoelectric, where the $\E$ and $\H$ couplings are accounted for in a symmetric way.
As in the ordinary case, the dynamical matrix is a nonanalytic function of degree zero in $\q$, and is exact in the harmonic approximation. For the sake of completeness, we address other issues, and in particular we solve a problem which might arise in first-principle implementations, where---differently than here---the basic fields are  $\E$ and $\B$ (not $\H$).
\end{abstract}

\pacs{63.20.-e,  75.85.+t, 63.20.dk}

\maketitle

\section{Introduction}

The zone-center dynamical matrix in polar dielectrics is comprised of an analytic term and a nonanalytic term; the latter accounts for the long range of Coulomb interactions, or equivalently for the coupling of macroscopic electric fields with the lattice. In high symmetry situations the nonanalytic term is responsible for the familiar longitudinal-transverse splitting of zone-center optical modes. The explicit form for the zone-center dynamical matrix in crystals of arbitrary symmetry was first provided in 1962 by Cochran and Cowley.\cite{Cochran62} Their phenomenological formula is exact within the harmonic approximation
and has been implemented much later in some first-principle codes.\cite{Giannozzi91,quantum,abinit} 

Magnetoelectrics (MEs) are insulators where electric fields control magnetization, and conversely magnetic fields control polarization; they attracted considerable theoretical and technological interest in recent times.\cite{Fiebig05,Eerenstein06,Iniguez08,Hehl08,Hehl09,Essin09,Wojdel09,Essin10,Wojdel10,Lee10,rap145,Bousquet11} In such materials both fields (electric and magnetic) are coupled to the lattice: therefore both contribute to the nonanalytic term and (in high symmetry cases) to the longitudinal-transverse splitting. In Ref. \onlinecite{rap145} it is shown that the time-honored Lyddane-Sachs-Teller relationship,\cite{Lyddane41,Kurosawa61,Lax71,Gonze97b} which applies to ordinary dielectrics only,  generalizes in a perspicuous way to MEs. In the present paper we provide the exact form of the zone-center dynamical matrix in a crystal of arbitrary symmetry, where the nonanalytic term accounts for the coupling of both fields to the lattice in a very symmetric way.

So far, we have not specified which macroscopic fields. In lattice dynamics, the natural choice is the pair $(\E,\H)$. In fact, they are both longitudinal, while the fields $\D$ and $\B$ are both transverse.\cite{Landau,rap_a30} Because of this key feature the contribution of $(\E,\H)$ to the restoring forces appears as a nonanalytic term, while the analytic one corresponds to setting $\E=\H=0$.
In contrast to this situation, the natural choice in the framework of first-principle calculations is the pair $(\E,\B)$; in fact a second order expansion of the energy per cell with the ordinary periodic boundary conditions correspond to $\E=\B=0$ and, therefore, {\it does not} provide as such the analytic term in the force-constant  matrix. This problem is addressed and its solution given.

Another issue addressed in the present work is the microscopic nature of the coupling of magnetic fields to the lattice, which at first may appear as counterintuitive. In fact, a microscopic magnetic field does not exert any force on a nucleus at rest. 

The plan of the paper is as follows. In Sec \ref{sec:diel} we establish our notations and we review the Cochran-Cowley formula for ordinary dielectrics.\cite{Cochran62,Giannozzi91} in Sec. \ref{sec:MEs} we show how this formula generalizes to MEs, arriving at our major result, \equ{central}. We then deal with other related issues. In Sec \ref{sec:LST} we discuss the relationship of the present work to a recent generalization (to MEs) of the Lyddane-Sachs-Teller relationship.\cite{rap145} In Sec \ref{sec:micro} we show how a magnetic field in MEs does indeed exert a force on a nucleus at rest. In Sec \ref{sec:EB} we address the $\H$ versus $\B$ issue, crucial for any first-principle implementation of the present result. Finally, in Sec. \ref{sec:conclu}
we draw our conclusions.

\section{Ordinary dielectrics} \label{sec:diel}

For the sake of completeness, as well as for establishing our notations, we provide here a derivation of the Cochran-Cowley\cite{Cochran62} phenomenological formula for the zone-center dynamical matrix in ordinary dielectrics of arbitrary symmetry.
 Whenever the crystal is polar and insulating, the dynamical matrix shows a nonanalytic behavior at the zone center, which accounts for field-lattice coupling. Only electrical fields are considered in this Section. 
 
\subsection{Notations}

We use compact notations leaving the Cartesian indices implicit throughout.
The electronic (called also clamped-nuclei or ``static high frequency'') dielectric tensor\cite{AM} is real symmetric and indicated as $\ee$; the zone-center analytic part of the force-constant matrix, indicated as $C_{ss'}$ is real symmetric for a simultaneous exchange of {\it both} its
(implicit) Cartesian indices {\it and} its basis indices $s,s'$. This analytic part  yields by  definition the second order expansion of the energy in the lattice-periodical displacements $\u_s$ at $\E=0$.  Whenever the force-constant matrix is computed from first principles, the ordinary choice of periodic boundary conditions is equivalent to assume $\E=0$; the computation provides then $C_{ss'}$ directly. The magnetic analogue is different in this respect: a discussion is provided in Sec. \ref{sec:EB}. The Born-charge tensors $Z^*_s$ and $Z^{*\dagger}_s$ (transpose) are nonsymmetrical Cartesian tensors. We further define the unit vectors in the $\q$-direction as
\[  \hq = \left(\begin{array}{c} \hat{q}_x \\  \hat{q}_y  \\ \hat{q}_z  \end{array} \right) , \qquad
\hq^\dagger = \left(\begin{array}{ccc}  \hat{q}_x  & \hat{q}_y  & \hat{q}_z  \end{array} \right) .
\] Therefore the norm is $\hq^\dagger \hq = 1$, while the dyadic product $\hq\,  \hq^\dagger$ is the projector in the direction of $\q$.

\subsection{$\D$ and $\E$ fields}

In presence of a long wavelength phonon of wavevector $\q$, the solid is macroscopically homogeneous in the plane normal to $\q$, while all macroscopic properties display a modulation in the direction of $\q$. It is immediate to realize that the component of $\D(\q)$ parallel to $\q$ and the component of $\E(\q)$ normal to $\q$ both vanish:\cite{Landau,rap_a30}  \[  \hq^\dagger \D(\q) = 0, \qquad (1 -  \hq \, \hq^\dagger) \E(\q) = 0 . \label{lt} \] Whenever nonvanishing, both $\D(\q)$ and $\E(\q)$ are nonanalytic functions of order zero in $\q$. In the following equations, it is tacitly assumed that only the leading term in $\q$ is considered. 

If $\P(\q)$ is the macroscopic polarization due to a phonon in zero $\E$ field (transverse polarization),
the fields $\D$ and $\E$ are related by \[ \D(\q) = \ee \E(\q) + 4 \pi \P(\q) \label{basic} . \] We now exploit \equ{lt} as follows: \[ 0 = \q^\dagger \D(\q) = \q^\dagger \ee \E(\q) + 4 \pi \q^\dagger \P(\q) \]
\[ \E(\q)  = \hq \, \hq^\dagger \E(\q) . \] From these it easily follows that \bea 0 &=& \hq^\dagger \ee \hq \, \hq^\dagger \E(\q) + 4 \pi \hq^\dagger \P(\q) \\  \E(\q) &=& - \frac{4\pi}{ \hq^\dagger \ee \hq} \hq \, \hq^\dagger \P(\q) , \label{dp} \eea which can be interpreted as the depolarization field for an arbitrary $\hq$-direction.

\subsection{Equations of motion}

The equations of motion in the harmonic approximation are \[ - M_s \omega^2(\q)  \u_s(\q) = f_s(\q), \label{eom0} \] where the forces and the fields, to leading order in $\q$, are
\bea  f_s(\q) &=& - \sum_{s'} C_{s s'} \u_{s'}(\q) + Z^{*\dagger}_s \E(\q) \nn \D(\q) &=&  \frac{4\pi}{\Omega} \sum_s Z^{*}_s \u_s(\q) + \ee \E(\q) . \label{eom1}
\eea In \eqs{eom0}{eom1} $\u_s$ are sublattice displacements, $M_s$ the corresponding nuclear masses, and $\Omega$ is the cell volume. From the second line of \equ{eom1} it is clear that the phonon polarization in zero $\E$ field is $\P(\q) = \Omega^{-1} \sum_s Z^{*}_s \u_s(\q)$, and the corresponding depolarization field at arbitray $\hq$ is then \[ \E(\q) = - \frac{4\pi}{ \Omega \; \hq^\dagger  \ee \hq} \hq \sum_s  \hq^\dagger  Z^{*}_s \u_s(\q) \label{depol} \] The first line of \equ{eom1} yields therefore
\[  f_s(\q) =  - \sum_{s'} \left[ C_{s s'}  +  \frac{4\pi}{\Omega} \frac { (Z^{*\dagger}_s \q )\; ( \q^\dagger   Z^{*}_{s'} ) }{  \q^\dagger  \ee \q} \right] \u_{s'}(\q) . \label{liter} \]
The quantitiy in parenthesis is indeed the usual expression for the force-constant matrix at the zone center, including the nonanalytic term, first obtained in 1962 by Cochran and Cowley\cite{Cochran62} and implemented much later in some first-principle codes.\cite{Giannozzi91,quantum,abinit} This confirms that  the matrix elements  at the zone center are indeed nonanalytic functions, homogeneous of degree zero in $\q$; we also remind that \equ{liter} applies to crystals of any symmetry. 

It is also expedient to provide an expression equivalent to \equ{liter} by restoring in it the normalized wavectors and indicating with ${\cal P}(\hq) = \hq \hq^\dagger$ the projector in the $\q$-direction. Then
\[  f_s(\q) =  - \sum_{s'} \left[ C_{s s'}  +  \frac{4\pi}{\Omega} \frac {Z^{*\dagger}_s {\cal P}(\hq)   Z^{*}_{s'}  }{  \hq^\dagger  \ee \hq} \right] \u_{s'}(\hq) . \label{liter2} \] It is immediate to verify in either \equ{liter} or \equ{liter2} that even the nonanalytic term is symmetric for a simultaneous exchange of both
the Cartesian indices and the  basis indices.

The nonanalytic term in the force constants accounts for an additional restoring force due to the depolarizing field; when the corresponding dynamical matrix is evaluated, this term can be interpreted as a generalized (squared) plasma frequency. We also notice that, at variance with the familiar high symmetry cases, in a low symmetry crystal the two terms (analytic and nonanalytic) in the dynamical matrix do not need to commute. When this is the case,  {\it all} zone-center modes are coupled to a nonvanishing $\E$ field (i.e. they are infrared active) while the $C_{s s'}$ by themselves do not correspond to any physical mode.

\section{Magnetoelectrics} \label{sec:MEs}

In any linear ME the role of the $3\times 3$ Cartesian tensor $\ee$ is played by a $6\times6$ response matrix---called ${\cal R}$ here---which yields the macroscopic fields $(\D,\B)$ in terms of $(\E,\H)$. We define the purely electronic (clamped-nuclei) response as\cite{rap145}
 \[ \left(\begin{array}{c} \D \\ \B \end{array} \right) = \rr \left(\begin{array}{c} \E \\ \H \end{array} \right) \equiv  \left(\begin{array}{cc} \ee & \aa \\  \aa^\dagger  & \mm \end{array} \right)  \left(\begin{array}{c} \E \\ \H \end{array} \right) , \label{resp} \] where $\mm$, and $\aa$ are  the clamped-nuclei  magnetic permeability and ME coupling tensor, respectively.
 
 \subsection{Depolarization and demagnetization fields}
 
If $\P(\q)$ and $\M(\q)$ are the macroscopic polarization and magnetization due to a phonon at $\E = \H =0$, the analogue of \equ{basic} is
 \[ \left(\begin{array}{c} \D(\q) \\ \B(\q) \end{array} \right) = \rr \left(\begin{array}{c} \E(\q) \\ \H(\q) \end{array} \right) +  4 \pi \left(\begin{array}{c} \P(\q) \\ \M(\q) \end{array} \right) . \label{basic2} \] We then exploit \equ{lt} together with its magnetic analogue \[ \hq^\dagger \B(\q) = 0, \qquad (1 -  \hq \, \hq^\dagger) \H(\q) = 0 , \label{ltm} \] to obtain the two scalar equations \bea 0 &=& \hq^\dagger \ee \hq \, \hq^\dagger \E(\q) + \hq^\dagger \aa  \hq \, \hq^\dagger \H(\q) + 4 \pi \hq^\dagger \P(\q)  \\ 0 &=&  \hq^\dagger \aa^\dagger \hq \, \hq^\dagger \E(\q) + \hq^\dagger \mm  \hq \, \hq^\dagger \H(\q) + 4 \pi \hq^\dagger \M(\q) \nonumber .
 \eea The scalars $\hq^\dagger \E(\q)$ and $\hq^\dagger \H(\q)$ are easily obtained by inverting the $2 \times 2$ matrix \[ {\cal M}(\hq) = \left(\begin{array}{cc} \hq^\dagger \ee \hq & \hq^\dagger \aa \hq \\  \hq^\dagger \aa^\dagger \hq & \hq^\dagger \mm \hq \end{array} \right) \label{M} \] 
 \[ \left(\begin{array}{c} \hq^\dagger \E(\q) \\ \hq^\dagger \H(\q) \end{array} \right) = -4 \pi {\cal M}^{-1}(\hq)  \left(\begin{array}{c} \hq^\dagger \P(\q) \\ \hq^\dagger \M(\q) \end{array} \right)  .  \] Finally we exploit once more $\E(\hq) =
 \hq \, \hq^\dagger \E(\q)$, $\H(\hq) = \hq \, \hq^\dagger \H(\q)$, to obtain both depolarization and demagnetization fields as  \[ \left(\begin{array}{c}  \E(\q) \\  \H(\q) \end{array} \right) = -4 \pi  \; {\cal M}^{-1}(\hq)  \left(\begin{array}{c}  \hq\hq^\dagger \P(\q) \\ \hq \hq^\dagger \M(\q) \end{array} \right)  .  \label{dpm} \] Both fields are longitudinal (parallel to $\hq$); \equ{dpm} clearly generalizes \equ{dp} to the ME case, and coincides with it for a magnetically inert material.
 A comment about notations: in order not to overburden notations, we indicate with ${\cal M}^{-1}(\hq)$ in \equ{dpm} the  $6 \times 6$ matrix, diagonal over its Cartesian indices, whose three $2 \times 2$ blocks are indeed the inverse of ${\cal M}(\hq)$ in \equ{M}.
 
 Not surprisingly, the fields $\E(\q)$ and $\H(\q)$ for a given $\hq$ direction depend on both $\P(\q)$ and $\M(\q)$, defined as the macroscopic polarization and magnetization in zero fields. This, indeed, is the hallmark of ME materials
 
 \subsection{Equations of motion}
 
One starts from the most general expression for the free energy $F(\{\u_s\},\E,\H)$, where the sublattice displacements and the fields $\E,\H$ are chosen as independent variables.\cite{Iniguez08,rap145}  The coefficients of its second order expansion, entering the equation of motion, are tensorial material constants which in general are all independent from each other. The second derivatives at zero displacements are proportional to $\rr$, while  the mixed second derivatives of $F$ with respect to the displacements and either $\E$ or $\H$ are the lattice coupling tensors (electrical and magnetic, respectively).

In the present notations the forces and the fields, to leading order in $\q$, are \[  f_s(\q) = - \sum_{s'} C_{s s'} \u_{s'}(\q) + Z^{*\dagger}_s \E(\q) + \zeta^{*\dagger}_s \H(\q)  \label{eom2} \]  \vspace{-0.5cm} $$ \left(\begin{array}{c} \D(\q) \\ \B(\q) \end{array} \right)  = \rr \left(\begin{array}{c} \E(\q) \\ \H(\q) \end{array} \right)  +  \frac{4\pi}{\Omega} \sum_s \left(\begin{array}{c} Z^*_s \\ \zeta^*_s \end{array} \right) \u_s(\q) ,  $$ where $\zeta^*_s$, first introduced by 
  \`I\~niguez,~\cite{Iniguez08} is the magnetic analogue of the Born effective-charge tensor. Notice that in \equ{eom2} the analytical term in the force-constants matrix  $C_{ss'}$ coincides {\it by definition} with the full force-constants matrix in zero $\E$ and $\H$ fields (not $\B$). This feature deserves discussion, provided in Sec. \ref{sec:EB}.

The macroscopic fields associated to a long-wavelength phonon of wavevector $\q$ and lattice displacements $\u_s(\q)$ is therefore, according to \equ{dpm}, \[ \left(\begin{array}{c}  \E(\q) \\  \H(\q) \end{array} \right) = -\frac{4 \pi}{\Omega} \; {\cal M}^{-1}(\hq)  \sum_s \left(\begin{array}{c} \hq \hq^\dagger Z^*_s \\ \hq \hq^\dagger \zeta^*_s \end{array} \right) \u_s(\q).  \label{fields} \] We simplify a bit the notations, by indicating from now on with ${\cal P}(\hq)$ the ``double projector'', i.e. the  block-diagonal matrix \[
{\cal P}(\hq) = \left(\begin{array}{cc} \hq \hq^\dagger & 0 \\ 0  & \hq \hq^\dagger \end{array} \right) . \] We notice that ${\cal P}(\hq)$ is a $6 \times 6$ matrix, diagonal on the field indices, while we remind that ${\cal M}^{-1}(\hq)$, also  a $6 \times 6$ matrix, is instead diagonal on the Cartesian indices. Therefore the two matrices ${\cal P}(\hq)$ and ${\cal M}^{-1}(\hq)$ commute.
It is also expedient  to define the ME lattice-coupling matrix: \[ \zz_s^\dagger = ( Z_s^\dagger , \zeta_s^\dagger) . \] Notice that $\zz_s^\dagger$  is a $3 \times 6$ matrix (and $\zz_s$ is  $6 \times 3$).

In these compact notations \equ{fields} becomes  \[ \left(\begin{array}{c}  \E(\q) \\  \H(\q) \end{array} \right) = -\frac{4 \pi}{\Omega} \; {\cal M}^{-1}(\hq)  {\cal P}(\hq) \sum_s \zz^*_s   \u_s(\q).  \label{fields2} , \] and finally replacing into the first line of \equ{eom2} we get the generalized form of \eqs{liter}{liter2} as \[ f_s(\q) = - \sum_{s'} \left[ C_{s s'} + \frac{4 \pi}{\Omega} \; \zz_s^\dagger {\cal M}^{-1}(\hq)  {\cal P}(\hq)  \, \zz^*_{s'} \right] \u_{s'}(\q)  \label{central} . \] Owing to the fact that ${\cal P}(\hq)$ and ${\cal M}^{-1}(\hq)$ commute, both terms in \equ{central} are---as they must be---symmetric for a simultaneous exchange of both the Cartesian indices and the  basis indices. In a low-symmetry crystal, the two terms do not in general commute, as indeed in the purely electrical case. \equ{central} is the central result of this work; it is a consequence of the fact that, in ME crystals, both macroscopic fields $\E$ and $\H$ are coupled to long-wavelength modes, and both are therefore at the root of the nonanalytic term.

\section{Lyddane-Sachs-Teller relationship} \label{sec:LST}

We address in this Section only crystals whose symmetry is orthorombic or higher, in which case all crystalline tensors can be simultaneously diagonalized. This is e.g. the case for the paradigmatic ME crystal Cr$_2$O$_3$. If we choose $\q$ along a principal axis, then the zone-center optical modes are either longitudinal ($\u_s$ parallel to $\q$), or transverse ($\u_s$ normal to $\q$).

Let us consider first, both for simple dielectrics and  for MEs, the  analytic term only in the zone-center  dynamical matrix \[ D^{\rm (analytic)}_{ss'} = \frac{1}{\sqrt{M_s M_{s'}}} C_{ss'} . \] When the Cartesian axes coincide with the principal axes, this matrix factorizes; for a crystal with an $(N+1)$-atom basis, each of the blocks leads to one zero eigenvalue (acoustic mode) and $N$ nonzero eigenvalues, which we call $\omega^2_n$ and correspond to the (squared) optic frequencies of the system for a long wavelength-mode with $\q$ normal to the chosen principal axis. The restoring forces responsible for these modes do not have any contribution from the macroscopic fields (either $\E$ or $\H$).

If we consider instead long-wavelength modes whose $\q$ vector is parallel to the chosen axis, then the projector ${\cal P}(\hq)$ in both \equ{liter2} or \equ{central} act like the identity (either $1 \times 1$ or $2 \times 2$, respectively), and the dynamical matrix has (in general) different eigenvalues and eigenvectors from the previous case. We indicate these longitudinal eigenvalues as $\tilde{\omega}^2_n$.

In this work we have not yet addressed the response of the system to a genuinely static perturbation. In macroscopic fields (either $\E$ or $\H$) the
total polarization and magnetization are due to the response of both the electronic system and the lattice. 

For ordinary dielectrics, is it customary to indicate as $\varepsilon_\infty$ the electronic dielectric tensor and with $\varepsilon_0$ the genuinely static one, which includes the lattice contribution.\cite{AM} 
In an high-symmetry situation we may choose a principal axis  and consider the scalar $\varepsilon_0/\varepsilon_\infty \geq 1$. The Lyddane-Sachs-Teller (LST) relationship in its original form\cite{Lyddane41} applies to a binary crystal, where the number of optic modes is $N=1$; it states that $\varepsilon_0/\varepsilon_\infty = \tilde{\omega}^2_1/\omega^2_1$, where the modes are polarized along the same chosen axis. Whenever $N > 1$, the ratio $\varepsilon_0/\varepsilon_\infty$ is related to a function of the $\tilde{\omega}^2_n$ and $\omega^2_n$.
One of the expressions for this function has the form of the ratio of a weighted harmonic average of the $\tilde{\omega}^2_n$, over a weighted harmonic average of the ${\omega}^2_n$. Obviously this yields the original LST relationship for $N=1$.

The LST relationship---either its original or generalized form\cite{Kurosawa61,Cochran62,Lax71,Gonze97b}---is exact in the harmonic approximation.  The two members of the identity are greater than one in polar crystals, as a consequence of the same physical effect: the coupling of the lattice with macroscopic electric fields.

In MEs we may generalize \equ{resp} by including the lattice contribution
\[  {\cal R}_0 = \left(\begin{array}{cc} \varepsilon_0 & \alpha_0 \\  \alpha_0^\dagger  & \mu_0 \end{array} \right)  . \label{resp2} \] When projected over a principal axis, both $\rr$ and ${\cal R}_0$  become $2 \times 2$ matrices. It has been recently shown\cite{rap145} that a scalar function of these two matrices takes the role of $\varepsilon_0/\varepsilon_\infty$ in a generalized LST relationship and equates the ratio of a weighted harmonic average of the $\tilde{\omega}^2_n$, over a weighted harmonic average of the ${\omega}^2_n$. At the root of the need for a generalization of the LST relationship is the fact that both fields $\E$ and $\H$ are coupled to the lattice on equal footing.
The explicit form of the generalized LST relationship for ME crystals  is given in Ref. \onlinecite{rap145}, where the noncrystalline and/or anharmonic cases are also addressed.

\section{Microscopic origin of the field-lattice couplings} \label{sec:micro}

By definition, the Born-effective charge tensor $Z_s^{*\dagger}$ yields the force exerted on the $s$-th nucleus at equilibrium by a macroscopic field $\E$. This clearly appears from  the first line of \equ{eom1} or even \equ{eom2} with $\H = 0$.
Equivalently its transpose yields the macroscopic polarization induced by a sublattice displacement $\u_s$ at zero $\E$ field. According to the second line of \equ{eom1} such polarization is in fact given by $Z^{*}_s \u_s / \Omega$.

The coupling tensor $\zeta^*_s$ plays a similar role in the magnetic case. 
If we set $\E = 0$ in  \equ{eom2},  $\zeta^{*\dagger}_s$ in its first line yields the force exerted on the $s$-th nucleus at equilibrium by a macroscopic field $\H$, while $\zeta^{*}_s$ in its second line yields the macroscopic magnetization $\zeta^{*}_s \u_s / \Omega$ induced by a sublattice displacement $\u_s$ at $\E=\H=0$.

The dual view just presented owes to the fact that both tensors are the second mixed derivatives, with respect to either $\E$ or $\H$ and to sublattice displacements $\u_s$, of a free energy, where $\u_s$, $\E$, and $\H$ are the independent variables.\cite{Iniguez08,rap145}

While we have addressed {\it macroscopic} fields so far, we switch to {\it microscopic} fields next. The force $f_s$ on a nucleus of (bare) charge $Z_s$ at zero displacement is equal to $Z_s \E_s$, where $\E_s$ is the microscopic field at site $s$. Such statement is no longer correct in a pseudopotential framework; to keep things simple, we adopt an all-electron view throughout this Section. We remind that, by definition, the macroscopic field in the lattice-periodical case is the cell average of the microscopic one.

In ordinary dielectrics we have therefore $Z^{*\dagger}_s \E = Z_s \E_s$, or equivalently \[ \E_s = \frac{1}{Z_s} Z^{*\dagger}_s \E , \] meaning that the tensor $Z^{*\dagger}_s / Z_s$ yields the microscopic field at site $s$ as a linear function of the macroscopic field. Notice that, while $Z_s$ is always a positive integer, the Born tensors fulfill the acoustic sum rule, and their sum vanishes: in the diagonal case they are real numbers bearing either sign. 

In MEs the force $f_s$ on the $s$th-nucleus at equilibrium is in general nonzero even at $\E=0$, provided that $\H \neq 0$, as shown by the first line of \equ{eom2}. Given that a magnetic field does not exert any force on a charge at rest, one may wonder about the microscopic origin of this force and of the corresponding $\zeta^{*\dagger}_s$ coupling tensors. The explanation is that the force on a nucleus is always  given by $f_s = Z_s \E_s$, but the microscopic field $\E_s$ at the nuclear site is a linear function of {\it both} macroscopic fields $\E$ and $\H$. In particular, $\E_s$ is in general nonzero (and linear in $\H$) even when $\E=0$.

\section{First-principle calculation of the force-constant matrix: $\H$ versus $\B$} \label{sec:EB}

All of the above results---and in particular the expressions for the zone-center force constants, \eqs{liter}{central}---are {\it exact} in the harmonic approximation, and apply to either empirical models or first-principle calculations. Indeed, the first occurrence of \equ{liter}, due to Cochran and Cowley,\cite{Cochran62} predates first-principle calculations by almost three decades.

For many dielectric materials the dynamical matrix is nowadays routinely computed from first principles,\cite{Giannozzi91} and for instance is provided by some codes in the public domain.\cite{quantum,abinit} At the zone center, the force-constant matrix has the form given by \eqs{liter}{liter2}. Therein, the ingredients of the second (nonanalytic) term are the clamped-nuclei dielectric tensor $\ee$ and the Born effective charge tensors $Z^*_s$: both quantities are usually computed using a linear response algorithm. As for the analytic part of the force-constant matrix $C_{ss'}$, it is computed via linear response as well, but could also be computed via zone-center "frozen phonons''. In both cases, the $C_{ss'}$ are the coefficients of the second order expansion of the total energy at equilibrium, assuming the ordinary periodic boundary conditions for solving Schr\"odinger equation. In fact, it has already been stressed that setting periodic boundary conditions is equivalent to set the macroscopic electric field $\E$ equal to zero; the magnetic case behaves quite differently in this respect.

In MEs a second order expansion of the total energy in the sublattice displacements $\u_s$, performed with the usual periodic boundary conditions, {\it does not} provide the analytic part of the force-constant matrix $C_{ss'}$ appearing in \equ{central}. As emphasized above, such analytic part implies $\E=\H=0$, while the ordinary periodic boundary conditions imply $\E=\B=0$.
In fact the real microscopic fields, in principle measurable inside the material, are $\E^{\rm (micro)}(\r)$ and $\B^{\rm (micro)}(\r)$; their potentials (scalar and vector, respectively) appear in the Schr\"odinger equation. Assuming periodic boundary conditions is tantamount to set the macroscopic $\E$ and $\B$ fields equal to zero. The existing  linear response codes\cite{quantum,abinit} routinely evaluate the response to $\E$ at $\B=0$; as for the response to $\B$, the needed algorithms appeared very recently.\cite{Essin10,Gonze11}.

Given the above, it is clear that the pair $(\E,\B)$ is apparently more fundamental than $(\E,\H)$. For this and other reasons the latter pair has been deemed a ``bastard pair'' and an ``unholy pair''.\cite{Hehl08,Hehl09} Here instead we have good reasons to use $(\E,\H)$ as independent variables in the free energy as well as in the corresponding equations of motion, \equ{eom2}. The key point is that, as emphasized e.g. by Landau-Lifshitz,\cite{Landau} $\E$ and $\H$ are both longitudinal, while $\D$ and $\B$ are both transverse: see  \eqs{lt}{ltm}; see also Ref. \onlinecite{rap_a30}.

We need therefore to address the following issue. Suppose we use any electronic structure code which uses $\E$ and $\B$ as control parameters: either setting $\E=0$ and $\B=0$, or providing the linear response to $\E$ and $\B$. The quantities entering our main formula, \equ{central}, {\it are not} these provided directly by the code; we show in the following their mutual relationships.

\subsection{Analytic term}

Suppose that  periodic boundary conditions are assumed  to expand the total energy in the zone-center displacements $\u_s$, and suppose that $\tilde{C}_{ss'}$ are the second order coefficients obtained in this way. Which is the relationship between these $\tilde{C}_{ss'}$ and the analytic coefficients ${C}_{ss'}$? Setting $\E=0$ and $\B=0$ in \equ{eom2} at $\q = 0$ we get
 \bea  f_s &=& - \sum_{s'}\tilde{ C}_{s s'} \u_{s'}  \nn &=&  - \sum_{s'} C_{s s'} \u_{s'} + \zeta^{*\dagger}_s \H \label{eom3a} \eea \[ \left(\begin{array}{c} \D \\ 0 \end{array} \right) = \rr \left(\begin{array}{c} 0 \\ \H \end{array} \right)  +  \frac{4\pi}{\Omega} \sum_s \left(\begin{array}{c} Z^*_s \\ \zeta^*_s \end{array} \right) \u_s .  \label{eom3b} \] From the second line of \equ{eom3b} the actual $\H$ field is \[ \H = - \frac{4\pi}{\Omega} \mu_\infty^{-1} \sum_s \zeta^*_s \u_s , \] hence the sought for relationship between the ${ C}_{s s'}$ and the $\tilde{ C}_{s s'}$ is \[ { C}_{s s'} = \tilde{ C}_{s s'} - \frac{4\pi}{\Omega}  \zeta^{*\dagger}_{s} \mu_\infty^{-1}  \zeta^{*}_{s'} \; .\] The interpretation of this is pretty clear: the force-constant matrix $\tilde{ C}_{s s'} $, computed at zero $\B$, includes the restoring forces due to nonvanishing $\H$. This contribution must be discounted to get the analytic force-constant matrix, which by definition means $\H$ (not $\B$) equal to zero.
 
\subsection{Nonanalytic term} 

We rewrite here, for the sake of clarity, \equ{eom2} at zero sublattice displacements (the $\q$ dependence becomes irrelevant):  \bea  f_s &=&   Z^{*\dagger}_s \E + \zeta^{*\dagger}_s \H  \nn  \left(\begin{array}{c} \D \\ \B \end{array} \right)  &=& \left(\begin{array}{cc} \ee & \aa \\  \aa^\dagger  & \mm \end{array} \right)  \left(\begin{array}{c} \E \\ \H \end{array} \right)  .   \label{eom0} \eea  For ordinary dielectrics, the existing codes\cite{quantum,abinit} essentially evaluate the (Hellmann-Feynman) forces and the electronic macroscopic polarization $\P$ linearly induced by $\E$, thus providing the tensors $Z_s^*$ and $\ee$ as \[ Z^{*\dagger}_s = \frac{\partial f_s}{\partial \E}, \qquad \ee = 1 + 4\pi  \frac{\partial \P}{\partial \E} . \label{der} \] This does not apply as such to MEs, because the codes implicitly set $\B=0$, not $\H=0$.

More generally, one could envisage a code which provides the linear response to $\E$ and $\B$: the algorithms have been just developed.\cite{Essin10,Gonze11} The output quantities would be forces $f_s$, macroscopic polarization $\P$ and magnetization $\M$. As for the magnetization, it is comprised of a spin (Zeeman) contribution and an orbital contribution. While the former contribution is in principle straightforward,\cite{Bousquet11} orbital magnetization has been understood only relatively recently.\cite{rap130}

We recast the lowest part of \equ{eom0} as 
\[  \left(\begin{array}{c} \E + 4\pi \P \\ \B \end{array} \right)  = \left(\begin{array}{cc} \ee & \aa \\  \aa^\dagger  & \mm \end{array} \right)  \left(\begin{array}{c} \E \\ \B -4\pi \M \end{array} \right) . \] A linear response code would in principle provide $\partial \P/\partial \E$, $\partial \P \partial \B$, $\partial \M / \partial \E$, and $\partial \M / \partial \B $, in terms of which the entries in the response tensor $\rr$ could be obtained by solving a linear system. 

As said above, the codes existing so far\cite{quantum,abinit} only provide $\partial \P/\partial \E$ at $\B=0$; whenever the computation addresses a ME such response {\it is not} simply related to $\ee$, as in \equ{der}. A simple calculations shows that the relationship is instead \[1 + 4\pi  \frac{\partial \P}{\partial \E} = \ee - \aa \mm^{-1} \aa^\dagger . \label{der2} \] The correction, being quadratical in $\alpha$, is definitely very small.

Once the clamped-nuclei response is evaluated as outlined above, we may address the coupling tensors in \equ{eom0}. The field $\H$ and the forces are given by
\bea \H &=& \mm^{-1} \B -  \mm^{-1} \aa^\dagger \E  \nn f_s &=& (Z^{*\dagger}_s  -  \zeta^{*\dagger}_s \mm^{-1} \aa^\dagger ) \E + \zeta^{*\dagger}_s  \mm^{-1} \B  . \label{cnew} \eea
The last term has a simple meaning: $\partial f_s /\partial \B$ at $\E=0$  is the {\it longitudinal} magnetic coupling tensor (analogue of the Callen effective charge), not the transverse one (analogue of the Born effective charge). The two are related by $\mm^{-1}$ (by $\ee^{-1}$ in the electrical analogue).

The existing codes provide $\partial f_s / \partial \E$ at $\B=0$. In MEs this does not coincide with $Z^{*\dagger}_s$; the appropriate modification of \equ{der} is 
\[ Z^{*\dagger}_s = \frac{\partial f_s}{\partial \E} + \zeta^{*\dagger}_s \mm^{-1} \aa^\dagger \label{derf} . \] 

\section{Conclusions} \label{sec:conclu}

Both in ordinary dielectrics and in MEs the zone-center dynamical matrix is a nonanalytic function of the wavevector $\q$, homogeneous of degree zero in $\q$. We have generalized the well established formula for ordinary dielectrics\cite{Cochran62,Giannozzi91,Gonze97b} to the ME case, where the lattice is coupled to both electric and magnetic fields. Our main result is \equ{central}, where the coupling appears in a symmetric way. The formula is {\it exact} (within the harmonic approximation), and is rooted in the formal equivalence of electric and magnetic fields in their coupling to the lattice in MEs. 

However, the orders of magnitude of electric and magnetic phenomena in condensed matter are not the same. ME effects are notoriously small,\cite{Fiebig05} and the corrections to the standard formula for dielectrics---\eqs{liter}{liter2}---in most cases are expected to be small as well. 
In oxides the electronic dielectric constants $\ee$ are typically in the range 2-3, while $|\mm -1|$ is of the order $10^{-4}$, bar when close to a ferromagnetic transition. Even the magnetic lattice coupling tensors $\zeta^*_s$ are smaller than their electric counterpart 
 $Z^*_s$ by orders of magnitude, e.g. in the paradigmatic crystalline ME, Cr$_2$O$_3$.\cite{Iniguez08}
As for the electronic ME coupling, little is known; a pioneering study by Bousquet et al.\cite{Bousquet11} addressed the Zeeman contribution to $\aa$, which is of the order $10^{-4}$ in Cr$_2$O$_3$ (Gaussian units are adopted here). More perspicuous effects are expected in nonconventional materials,\cite{Fiebig05} such as those where the ME effect can be tuned.\cite{Lee10}

Besides the major result of this work, \equ{central} we have also discussed other issues: (i) The relationship of this work to the Lyddane-Sachs-Teller relationship for MEs, recently published;\cite{rap145} (ii) The microscopic origin of the coupling of magnetic fields to the lattice, which may look counterintuitive; (iii) The relationship to first-principle implementations, where in the simplest cases $\E$ and $\B$ (not $\H$) are the control parameters for solving Schr\"odinger equation.

\section*{Acknowledgments} I thank the anonymous Referee who prompted me to write Sec. VI B, originally missing.
Work partially supported by the ONR Grant N00014-11-1-0145.


\end{document}